\documentclass[aps,prb,twocolumn,groupedaddress]{revtex4}
\usepackage{amsmath}
\usepackage{amssymb}
\usepackage{graphicx}

\begin{document}

\title{Anisotropic current-induced spin accumulation \\ in the
two-dimensional electron gas with spin-orbit coupling}

\author{Maxim Trushin and John Schliemann}

\affiliation{Institut f\"ur Theoretische Physik, Universit\"at Regensburg,
D-93040 Regensburg, Germany}

\date{\today}

\begin{abstract}
We investigate the magnetoelectric (or inverse spin-galvanic) effect in the
two dimensional electron gases with both Rashba and Dresselhaus 
spin-orbit coupling using an exact solution of the Boltzmann equation
for electron spin and momentum. The spin response to an in-plane
electric field turns out to be highly anisotropic while the usual charge
conductivity remains isotropic, contrary to earlier statements.
\end{abstract}

\pacs{72.10.-d \sep 72.20.My \sep 72.25.Dc }

\keywords{spin accumulation, Rashba and Dresselhaus spin-orbit coupling}

\maketitle

\section{Introduction}

Spin phenomena in semiconductor structures lie at the very heart of the
emerging field of spintronics and are 
to a major and still growing direction of solid-state research. Among the 
plethora of concepts and ideas, the magneto-electric effect
in the two-dimensional electron gas (2DEG),
sometimes also referred to as the inverse spin-galvanic effect, has
attracted particular interest from an both experimental
\cite{Nature2002ganichev,MMM2006ganichev,APL2004silov,PRL2006yang,PRL2006stern} and theoretical \cite{SSC1990edelstein,JETP1991aronov,PRB2003inoue,PRB2006huang,PRB2006jiang} point of view. This effect amounts in spin accumulation as
a response to an applied in-plane electric field and is therefore a possible
key ingredient towards all-electrical spin control in semiconductor structures,
a major goals in today's spintronics research.

We find it important to emphasize
that the phenomenon studied below is {\em not} the intrinsic spin Hall effect
\cite{Science2003murakami,PRL2004sinova} though it would manifest in a very similar
manner. The mechanisms involved in these two phenomena are, however, very different.
In particular, the momentum relaxation due to the impurity scattering
is necessary for the spin accumulation investigated in present paper, contrary
to the intrinsic spin Hall effect where ballistic regime is preferred
\cite{Science2003murakami,PRL2004sinova}.

In this paper we investigate the magnetoelectric effect
in 2DEG's with both Rashba and Dresselhaus spin-orbit coupling.
Our study is based on an exact analytical solution to the Boltzmann
equation for electron spin and momentum in the presence of
$\delta$-function shape static impurities.
Regarding the spin degree of freedom,
our solution neglects off-diagonal elements of the semiclassical distribution
matrix in the eigenbasis (or helicity basis) of the single-particle 
Hamiltonian, an approximation which is shown to be valid at sufficiently
high temperatures common in experiments. 

As a result, the magnetoelectrical effect, i.e. the magnetic response to 
an in-plane electric field, turns out to be highly anisotropic, whereas the
usual charge anisotropy remains isotropic, contrary to earlier statements
\cite{PRB2003schliemann}.

The paper is organized as follows.
First, we present an analytical solution of the semiclassical spin-incoherent
Boltzmann equation. Second, we solve the spin-coherent kinetic equation.
Finally, we discuss the plausibility of solutions obtained
and apply them for investigations of the electric current and spin accumulation.

\section{Solution of the kinetic equation}

We consider the Hamiltonian as a sum of the kinetic energy
and two spin-orbit coupling terms:
Rashba \cite{JETPL1984bychkov} and Dresselhaus \cite{PRB1955dresselhaus}.
Then, the Hamiltonian takes the form
\begin{equation}
H=\frac{\hbar^2(k_x^2+k_y^2)}{2m}+\alpha(\sigma_x k_y-\sigma_y k_x)
+\beta(\sigma_x k_x-\sigma_y k_y).
\end{equation}
Here, $\sigma_{x,y}$ are the Pauli matrices,
$k_{x,y}$ are the electron wave vectors, and  $m$ is the effective electron mass.
Introducing the angle $\gamma_k$ so, that
\begin{equation}
\tan\gamma_k=\frac{\alpha k_x + \beta k_y}{\beta k_x + \alpha k_y},
\end{equation}
we obtain the following spinors as the eigen functions of the Hamiltonian $H$
\begin{equation}
\label{psi}
\Psi_{\pm}(x,y)=\frac{1}{\sqrt{2}}{\mathrm e}^{ik_x x+ik_y y}\left(\begin{array}{c}
1 \\ 
\pm {\mathrm e}^{-i\gamma_k}
\end{array} \right).
\end{equation}
The energy spectrum has the form
\begin{equation}
\label{spectrum}
E_\pm=\frac{\hbar^2 k^2}{2m}
\pm\sqrt{(\alpha k_x + \beta k_y)^2+(\beta k_x + \alpha k_y)^2},
\end{equation}
where $k=\sqrt{k_x^2+k_y^2}$.
The velocity matrix in the helicity basis (\ref{psi}) is not diagonal,
and its elements read
\begin{eqnarray}
\label{v11x}&&
v_x^{11(22)}=
\hbar k_x/m \pm (\beta\cos\gamma_k + \alpha\sin\gamma_k)/\hbar, \\
\label{v12x}&&
v_x^{12(21)}=\pm i(\beta\sin\gamma_k - \alpha\cos\gamma_k)/\hbar;\\
\label{v11y}&&
v_y^{11(22)}=
\hbar k_y/m \pm (\alpha\cos\gamma_k + \beta\sin\gamma_k)/\hbar, \\
\label{v12y} && 
v_y^{12(21)}=\pm i(\alpha\sin\gamma_k - \beta\cos\gamma_k)/\hbar.
\end{eqnarray}
For the diagonal elements of the velocity matrix we also use
simplified notations $\mathbf{v^{11(22)}}\equiv\mathbf{v_\pm}$.

In the following, it will be convenient to use polar coordinates
$k_x=k\cos\theta$, $k_y=k\sin\theta$. Then, the spectrum read
$E_\pm=\hbar^2 k^2/(2m)\pm\mid k \mid \kappa_\theta$,
where
$$
\kappa_\theta=\sqrt{\alpha^2+\beta^2+4\alpha\beta\sin\theta\cos\theta}
$$
is the generalized spin-orbit interaction constant for a given direction
of motion. The wave vectors for a given energy read
\begin{equation}
\label{kv}
k_\pm=\mp\frac{m}{\hbar^2}\kappa_\theta
+\sqrt{\left(\frac{m}{\hbar^2}\right)^2\kappa^2_\theta+\frac{2mE}{\hbar^2}},
\end{equation}
and the expression for $\gamma_k$ takes the form
$$\tan\gamma_k=\frac{\alpha \cos\theta + \beta \sin\theta}{\beta \cos\theta + \alpha \sin\theta}.$$

We model the influence of impurities using
$\delta$-potential scattering $V=\zeta\delta(x,y)$.
The scattering probability between the states
with $k,s$ and $k',s'$ then read
\begin{equation}
\label{prob}
w(ks;k's')=\frac{\pi\hbar^2}{m \tau}
\left[1+ss'\cos(\gamma_k-\gamma_{k'})\right]
\delta(E_{sk}-E_{s'k'}),
\end{equation}
where $\tau$ is the relaxation time
which relates to $\zeta$ as $\tau=\hbar^3/(m\zeta^2)$.
Note, that the spin-dependent factor in Eq.~(\ref{prob}) does not depend on
the particular form of $V(x,y)$ in the Born approximation.
We expect, therefore, that such a simple scattering model as
$\delta$-function shape static impurities is quite reliable for the description
of spin-dependent transport phenomena.

\subsection{Spin-incoherent kinetic equation}

In general, the equation has the form
\begin{equation}
\label{master}
\frac{\partial f_s(\mathbf{k})}{\partial t}+\mathbf{v_s}\frac{\partial f_s(\mathbf{k})}{\partial \mathbf{r}}
+(-e\mathbf{E})\frac{\partial f_s(\mathbf{k})}{\hbar \partial \mathbf{k}} =
\left(\frac{\partial f_s}{\partial t}\right)_\mathrm{coll}.
\end{equation}
Here, $s=\pm$ is the spin index, and $\mathbf{v_s}$ are the diagonal elements
(\ref{v11x}) and (\ref{v11y}) of the velocity matrices.

To solve Eq.~(\ref{master}) we follow the standard procedure
widely spread in literature (see Ref.~\cite{abrikosov1988}, or any textbook on solid state physics).
We write down the distribution function as
$f_s=f_s^0+f_s^1+f_s^2$, where $f_s^0$ is the Fermi
distribution and $f_s^{1,2}\ll f_s^0$.
Then, at zero temperature gradient and constant electric field
in the linear responce regime Eq.~(\ref{master}) takes the form
\begin{equation}
\label{master2}
-e\mathbf{E}\mathbf{v_s}\left[-\frac{\partial f^0(E_{sk})}{\partial E_{sk} }\right]=
\mathrm{St}[f_s(\mathbf{k})].
\end{equation}
Assuming elastic scattering fullfilling the microreversibility condition the scattering
operator can be written as
\begin{equation}
\label{st}
\mathrm{St}[f_s(\mathbf{k})]=\sum\limits_{s'}\int\frac{d^2k'}{(2\pi)^2}
\left\{w(\mathbf{k}s;\mathbf{k'}s')[f_s^1(\mathbf{k})+f_s^2(\mathbf{k})-\right.
$$
$$
\left.
-f_{s'}^1(\mathbf{k'})-f_{s'}^2(\mathbf{k'})]\right\},
\end{equation}
where $w(\mathbf{k}s;\mathbf{k'}s')$ is given by Eq.~(\ref{prob}).

We would like to emphasize that besides abovementioned conventional
assumptions our solution is exact.
It means that in contrast to a number of previous works
we {\em do not} make any further approximation from this point.

The solution of Eq.~(\ref{master2}) has the form
\begin{eqnarray}
\label{solution1}
&&
f^1_s=-\tau e\mathbf{E}\mathbf{v_s}
\left[-\frac{\partial f^0(E_{sk})}{\partial E_{sk} }\right],\\
\label{solution2}
&&
f^2_s=-s\frac{\tau e \mathbf{E}}{\hbar\kappa_\theta}\left[-\frac{\partial f^0(E_{sk})}{\partial E_{sk} }\right] \times\\
\nonumber
&&
\times\left[(a_x\cos\theta+b_x\sin\theta){\mathbf e_x}+
(a_y\cos\theta+b_y\sin\theta){\mathbf e_y}\right].
\end{eqnarray}
Here, $f^1_s$ is the conventional solution of the Boltzmann equation,
whereas an addition $f^2_s$ is usually discarded since it
is much smaller than $f^1_s$.
However, this anisotropic addition plays an important role
in the spin accumulation mechanism as it is shown below.

In order to define the unknown coefficients $a_{x,y}$, $b_{x,y}$,
we substitute $f^{1,2}_s$ into Eq.~(\ref{master2})
(see Appendix for details)
and get the following equations
\begin{eqnarray}
\nonumber
&& a_x\frac{\alpha^2+\beta^2-\mid\alpha^2-\beta^2\mid}{4\alpha\beta}-\frac{b_x}{2}=\\
\label{a} &&
\alpha\beta
\left[1+\frac{(\alpha^2+\beta^2)^2}{4\alpha^2\beta^2}\left(
\frac{\mid\alpha^2-\beta^2\mid}{\alpha^2+\beta^2}-1\right)\right],   \\
\label{b}
&& b_x\frac{\alpha^2+\beta^2-\mid\alpha^2-\beta^2\mid}{2\alpha\beta}-a_x=
\mid \alpha^2-\beta^2 \mid.
\end{eqnarray}
The equations for $a_{y}$ and $b_{y}$ can be obtained from
Eqs.~(\ref{a})--(\ref{b}) by the substitution $a_{x}\rightarrow b_y$,
$b_{x}\rightarrow a_y$. From these equations
one can easily find that $a_x=b_y=-(\alpha^2+\beta^2)$ and $a_y=b_x=-2\alpha\beta$.

Finally, we substitute (\ref{v11x}), (\ref{v11y}) into Eq.~(\ref{solution1}),
sum up $f_s^1$ and $f_s^2$,
and write down the solution of Eq.~(\ref{master2}) in the following elegant form
\begin{equation}
\label{elegant}
f_s=f_s^0+(-e\mathbf{E})\mathbf{k}\frac{\hbar\tau}{m} 
\left[-\frac{\partial f^0(E_{sk})}{\partial E_{sk}}\right].
\end{equation}
This solution should be compared with the expression in the absence of
spin-orbit coupling which can be formulated in terms of the velocity
as follows
\begin{equation}
\label{noSO}
f_s=f_s^0+(-e\mathbf{E})\mathbf{v}_s\tau 
\left[-\frac{\partial f^0(E_{sk})}{\partial E_{sk}}\right],
\end{equation}
where $\mathbf{v}_s=\hbar\mathbf{k}/m$ is the velocity.
Note that this latter relation holds only in the absence of spin-orbit coupling.

\subsection{Spin-coherent kinetic equation}

Since we deal with the constant electric field only,
the equation reads \cite{PRL2006khaetskii}
\begin{equation}
\label{d-h}
(-e{\mathbf E}) \frac{\partial \hat{f}({\mathbf k})}{\hbar \partial {\mathbf k}}+
\frac{i}{\hbar}\left[\hat{H},\hat{f}({\mathbf k})\right]=
\left(\frac{\partial \hat{f}}{\partial t}\right)_\mathrm{coll}.
\end{equation}
Following Ref.~\cite{PRL2006khaetskii}, we rewrite
Eq.~(\ref{d-h}) in the helicity basis where the Hamiltonian
is diagonal, and the equation takes the form
\begin{equation}
\label{d-h-1}
\frac{e{\mathbf E}}{\hbar}\frac{\partial}{\partial {\mathbf k}}\left(\begin{array}{cc}
f_{11} & f_{12} \\ 
f_{21} & f_{22}
\end{array}\right)+
\frac{i e{\mathbf E}}{2\hbar}\frac{\partial \gamma_k}{\partial {\mathbf k}}
\left(\begin{array}{cc}
f_{21}-f_{12} & f_{22}-f_{11} \\ 
f_{11}-f_{22} & f_{12}-f_{21}
\end{array}\right) 
$$
$$
+\frac{i}{\hbar}
\left(\begin{array}{cc}
0 & f_{12}(E_+ - E_-) \\ 
f_{21}(E_- - E_+) & 0
\end{array}\right)= {\mathrm St}[\hat{f}({\mathbf k})],
\end{equation}
where
\begin{equation}
\label{g-der}
\frac{\partial \gamma_k}{\partial {\mathbf k}}=
\frac{\alpha^2-\beta^2}{k\kappa_\theta^2}
\left({\mathbf e_x}\sin\theta-{\mathbf e_y}\cos\theta\right).
\end{equation}
The collision term reads \cite{PRL2006khaetskii}
\begin{eqnarray}
\nonumber
&&
{\mathrm St}[\hat{f}({\mathbf k})]_{ss_1}=
-\int\frac{d^2 k'}{(2\pi)^2}\sum\limits_{s',s'_1}
\{[\delta(E_{s'k'}-E_{sk}) \\
\nonumber &&
+\delta(E_{s'_1 k'}-E_{sk})]
K^{ss_1}_{s' s'_1}f_{s' s'_1}(k')
-\delta(E_{s'k}-E_{s'_1 k'}) \\
&&
\times [K^{ss'}_{s'_1 s'_1}f_{s' s_1}(k)+K^{s's_1}_{s'_1 s'_1}f_{s'_1 s'_1}(k)]\},
\end{eqnarray}
where
\begin{eqnarray}
\nonumber
&&
K_{s's'_1}^{ss_1}=\frac{\pi\hbar^2}{4 m \tau}
[1+ss's_1s'_1+ \\
\label{K}
&&
+ss'{\mathrm e}^{i(\gamma_k-\gamma_{k'})}
+s_1s'_1{\mathrm e}^{-i(\gamma_k-\gamma_{k'})}].
\end{eqnarray}
Eq.~(\ref{K}) was derived for $\delta$-potential scattering
introduced above, and each $K_{s's'_1}^{ss_1}$ contains either
$\sin(\gamma_k-\gamma_{k'})$ or $1\pm\cos(\gamma_k-\gamma_{k'})$
depending on the product of $ss's_1s'_1$.

In order to simplify Eq.~(\ref{d-h-1}), we assume
$k\kappa_\theta\ll T$, where $T$ is the temperature.
This assumption will be discussed later in details.
Now, we expand the Fermi distribution function
$f^0(E_{sk})$ in terms of $k\kappa_\theta / T$ so, that
Eq.~(\ref{d-h-1}) in the linear responce regime takes the form
\begin{widetext}
\begin{equation}
\label{d-h-2}
\left(\begin{array}{cc}
 e{\mathbf E}\mathbf{v^{11}}\frac{\partial f^0 (E_{+k})}{\partial E_{+k} } &
\frac{e{\mathbf E}}{2}\mathbf{v^{12}}
\left[\frac{\partial f^0(E_{+k})}{\partial E_{+k} }+\frac{\partial f^0(E_{-k})}{\partial E_{-k} }\right]
+\frac{i}{\hbar}f_{12}(E_+ - E_-) \\ 
\frac{e{\mathbf E}}{2}\mathbf{v^{21}}
\left[\frac{\partial f^0(E_{+k})}{\partial E_{+k} }+\frac{\partial f^0(E_{-k})}{\partial E_{-k} }\right]
+\frac{i}{\hbar}f_{21}(E_- - E_+)
 & e{\mathbf E}\mathbf{v^{22}}\frac{\partial f^0 (E_{-k})}{\partial E_{-k} }
\end{array}\right) = {\mathrm St}[\hat{f}({\mathbf k})].
\end{equation}
\end{widetext}
Though Eq.~(\ref{d-h-2}) is still quite cumbersome,
it is easy to check that $f_{12}=f_{21}=0$
and $f_{11(22)}$ given by Eq.~(\ref{elegant}) represent the solution.
The details of calculations can be found in Appendix.
Thus, the solution for the spin-incoherent Boltzmann equation
is the same as for the spin-coherent kinetic equation 
at high temperatures. This can be explained in what follows.

\section{Results and Discussion}

Note, that the off-diagonal elements of the distribution function have
essentially quantum mechanical origin since they correspond to
the off-diagonal elements of the density matrix.
Classically, an electron can be in only one state of two,
and, therefore, off-diagonal terms vanish here.
In contrast, in the spin-coherent kinetic equation the off-diagonal terms
can be essential. However, in the real samples
the quantum effects are negligible at room temperatures
because of the temperature smearing.
Indeed, the spin-orbit coupling
constant $\kappa_\theta$ is of the order of $10^{-11}\,\mathrm{eV\cdot m}$
for typical InAs samples.
To be specific, let us take n-type InAs quantum well containing
the 2DEG used for photocurrent measurements at room temperature \cite{PRL2004ganichev}.
The parameters are as follows:
$\alpha/\beta=2.15$, mobility is about
$2\cdot 10^{4}\, \mathrm{cm^{2}/(V\cdot s)}$
and free carrier density is $1.3\cdot 10^{12}\,\mathrm{cm^{-2}}$.
The latter allows us to estimate characteristic
Fermi wave vector  $k_F=\sqrt{2\pi n_e}\simeq 3\cdot 10^{6}\,\mathrm{cm^{-1}}$.
Thus, the spin-orbit splitting energy $k_F\kappa_\theta$ is about $3\,\mathrm{meV}$,
that is much smaller than $T_\mathrm{room}=25\,\mathrm{meV}$, and
our solution is suitable for description of a large variety of experiments.

To study the spin accumulation we calculate
the net spin density, whose $x,y$-components read
\begin{equation}
\label{sd}
\langle S_{x,y} \rangle=\sum\limits_s\int\frac{d^2 k}{(2\pi)^2}
S_{x,y}(k,s) f_s(\mathbf{k}),
\end{equation}
where $S_x(k,s)=\frac{s}{2}\cos\gamma_k$, $S_y=-\frac{s}{2}\sin\gamma_k$
are the spin expectation values.
(We do not consider $S_z$-component since it is zero.)

The integral over ${\mathbf k}$ can be taken easily
making the substitution $\varepsilon=E(s,k)$ and assuming that
$-\partial f^0(\varepsilon)/\partial \varepsilon=
\delta(E_F-\varepsilon)$.
This assumption is reasonable with respect to the system studied in Ref.~\cite{PRL2004ganichev}
from which one can deduce a Fermi energy of the order of $100\,\mathrm{meV}$,
which is clearly larger than room temperature.
Since our solution is valid for temperatures much higher
than the spin-orbit splitting energy $k_F\kappa_\theta$,
the inequality describing the applicability of our results obtained below reads
$$
k_F \kappa_\theta \ll T \ll E_F.
$$

\begin{figure}
\includegraphics{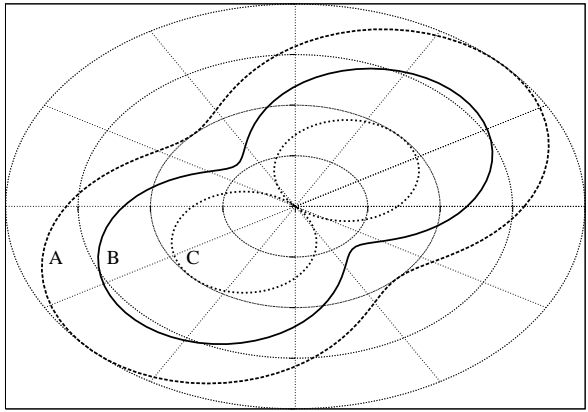}
\caption{\label{fig} Spin accumulation (in arbitrary units) vs. direction
of the electric field in polar coordinates for different Rashba and Dresselhaus constants:
A --- $\alpha=3\beta$, B --- $\alpha=2.15\beta$, C --- $\alpha=\beta$.
The curve B corresponds to the real situation in n-type InAs \cite{PRL2004ganichev}.}
\end{figure}

The rest integrals over the polar angle can be taken analytically.
After some algebra we have
\begin{equation}
\label{tensor}
\langle \mathbf{S} \rangle = \frac{em\tau}{2\pi\hbar^3}
\left(\begin{array}{cc} \beta & \alpha \\  -\alpha & -\beta \end{array}\right)
\mathbf{E}.
\end{equation}
Then, the magnitude of the spin accumulation
$\langle S \rangle=\sqrt{\langle S_{x}\rangle^2 + \langle S_{y}\rangle^2}$
is given by
\begin{equation}
\label{s_aver}
\langle S \rangle=\frac{eEm\tau}{2\pi \hbar^3}
\sqrt{\alpha^2+\beta^2+2\alpha\beta\sin(2\widehat{\mathbf{E}\mathbf{e}}_x)}.
\end{equation}
It is interesting to note, that $\langle S \rangle$ depends on the direction of the
electric field (see Fig.~\ref{fig}), i. e. the spin accumulation is anisotropic.
To our knowledge, this interesting feature was not noticed in the literature so far.
If $\beta=0$ then $\langle S \rangle=eEm\tau\alpha/(2\pi \hbar^3)$
that is in agreement with Refs.~\cite{SSC1990edelstein,PRB2003inoue}.
If the electric field is applied along the $x$-direction
(the case studied in Ref.~\cite{PRB2006huang}), then the spin density is
$\langle S \rangle=eEm\tau\sqrt{\alpha^2+\beta^2}/(2\pi \hbar^3)$.
This result contradicts to Ref.~\cite{PRB2006huang}, where
the spin density has some strange kinks as a function of
$m\alpha^2/\hbar^2$ and $m\beta^2/\hbar^2$.
However, one can see from Eq.~(\ref{s_aver}) that the kinks
in the dependencies of $\langle S \rangle$ on $\alpha$ (or $\beta$) could not take place.

Relying on the effect depicted in Fig.~\ref{fig}
the following novel spintronic device can be proposed.
Let us attach two pairs of contacts to the 2DEG so that
the first pair provides the electric current along the
crystallographic axis corresponding to the minimal spin accumulation
(i. e. $[1\overline{1}0]$-axis for $[001]$-grown InAs samples \cite{PRL2004ganichev}),
and the second one is connected along the perpendicular axis.
Then, the spin accumulation depends strongly
on which contacts the transport voltage is applied, and its anisotropic
contribution can be extracted easily using optical methods
\cite{APL2004silov,MMM2006ganichev,PRL2006yang,PRL2006stern}
or just measuring the magnetization.
To give an example, applying the electric field of $20\,\mathrm{V/m}$
(which corresponds to the current density of $1\, \mathrm{mA/cm}$)
we obtain the magnetization difference of the order
of $10^6\mu_B$ per $\mathrm{cm^2}$, where $\mu_B$ is the Bohr magneton.
This is comparable to the Pauli magnetization at the magnetic
fields of a few gauss.
It is also interesting to note the small characteristic switching
time $\tau\simeq 10^{-13}\,\mathrm{s}$ of the device proposed.
Therefore, besides the fundamental importance of 
such an experiment, our four terminal device could find
some applications as a high-speed spin switch.

At the end of the discussion, let us turn to the
charge current whose density can be found as
\begin{equation}
\label{current}
\mathbf{j}=-e\sum\limits_s\int\frac{d^2 k}{(2\pi)^2} \mathbf{v_s} f_s (\mathbf{k}).
\end{equation}
The integral over $\mathbf{k}$ in (\ref{current})
can be taken in the same manner as in (\ref{sd}),
and after some algebra the conductivity tensor $\sigma$ takes the form
\begin{equation}
\sigma_{yy(xx)}=\frac{e^2 \tau }{\pi\hbar^2}\left(
\frac{m(\alpha^2+\beta^2)}{\hbar^2} + E_F\right),
\end{equation}
and, most surprisingly, $\sigma_{xy(yx)}=0$.
It is convenient to express $\sigma_{yy(xx)}$
via the electron concentration (\ref{conc}) in a 2DEG with spin-orbit interactions
\begin{equation}
\label{conc}
n_e=\frac{m E_F}{\pi \hbar^2} + \left(\frac{m}{\hbar^2}\right)^2
\frac{\alpha^2+\beta^2}{\pi}.
\end{equation}
Then, the conductivity takes much simpler form, namely
$\sigma_{yy(xx)}=e^2 n_e \tau/m$.
This is the Drude formula, i. e. the conductivity is just a number
(not a tensor), though the distribution function (\ref{elegant}) is anisotropic.
This result is in contrast with the findings
of Ref.~\cite{PRB2003schliemann}, where a subtle approximation
generated off-diagonal elements in the conductivity tensor \cite{loss}.
This artifact is absent in our truly exact solution:
The electrical conductivity is isotropic for any relation between $\alpha$ and $\beta$.
The latter simplifies essentially the relation between the spin
and charge current densities, which takes the form
\begin{equation}
\label{s-j}
\langle \mathbf{S} \rangle = \frac{m^2}{2\pi\hbar^3 en_e}
\left(\begin{array}{cc} \beta & \alpha \\  -\alpha & -\beta \end{array}\right)
\mathbf{j}.
\end{equation}
We find it useful to write down Eq.~(\ref{s-j})
in the basis $\frac{1}{\sqrt{2}}(1;1)$, $\frac{1}{\sqrt{2}}(1;-1)$.
Then the relation between $\langle \mathbf{S} \rangle$ and 
$\mathbf{j}$ takes the simpler form
\begin{equation}
\label{s-j2}
\langle \mathbf{S} \rangle = \frac{m^2}{2\pi\hbar^3 en_e}
\left(\begin{array}{cc} 0 & \beta-\alpha \\  \beta+\alpha & 0 \end{array}\right)
\mathbf{j}.
\end{equation}
From this equation one can see easily, that
the spin accumulation is strongly anisotropic if
the constants $\alpha$ and $\beta$ are close to each other.
The reason of such an anisotropy is the angular dependence of the
dispersion law (\ref{spectrum}).
Indeed, the spin-orbit splitting is different for different direction
of the momentum. Therefore, the spin precession frequency
depends essentially on the direction of the electron motion.
This leads to the anisotropic spin relaxation times 
(see e. g. Ref.~\cite{PRB1999averkiev}),
and, thus, the anisotropy of the spin accumulation occurs.
In particular, the electric current of arbitrary strength applied
along the $[1\overline{1}0]$ crystallographic axis does not lead
to any spin accumulation at $\alpha=\beta$.
The latter is due to the vanishing spin splitting along
the $[1\overline{1}0]$-axis
(see e. g. Ref.~\cite{PRL2003schliemann}).

Eqs. (\ref{s-j}) and (\ref{s-j2}) are the main result of our work which
can be applied directly
to experimental studies of current-induced spin accumulation
in $[001]$-grown InAs samples.
The theory developed here is, of course, applicable to arbitrary
oriented heterostructures after a few minor changes regarding
the spin-orbit coupling terms in the Hamiltonian.

\section{Conclusions}

In summary, we solved 
the semiclassical Boltzmann equation analytically for 2DEGs with arbitrary large 
spin-orbit interactions of both Rashba and Dresselhaus type.
Moreover, we demonstrated that this solution is also suitable
for the spin-coherent case at sufficiently high temperatures
common to experiments.
Using this solution, we discovered the anisotropy
of the current-induced spin accumulation, though
the conductivity remains isotropic.
Finally, our analytical study is expected to be a reliable starting point
for further investigations of spin dependent electron transport.

We thank S.D. Ganichev and D. Loss for useful discussions and
acknowledge financial support from Collaborative Research Center 689.

\begin{appendix}
\begin{widetext}
\section{The collision integral calculations}

To take the integrals of the form $\int\frac{d^2k'}{(2\pi)^2}F(k',E_{s'k'})$
it is convenient to perform the substitution $E_{s'k'}=\varepsilon'$,
where $F(k',E_{s'k'})$ is a given function, and
$E_{s'k'}$ is the dispersion relation (\ref{spectrum}).

\subsection{Spin-incoherent case}

The first term of $\mathrm{St}[f_s(\mathbf{k})]$ given by
Eq.~(\ref{st}) reads
\begin{eqnarray}
\nonumber
&& \sum\limits_{s'}\int\frac{d^2k'}{(2\pi)^2}
w(\mathbf{k}s;\mathbf{k'}s')[f_s^1(\mathbf{k})+f_s^2(\mathbf{k})]=
\left[f_s^1(\mathbf{k})+f_s^2(\mathbf{k})\right]
\sum\limits_{s'}\int\limits_0^{2\pi}\frac{d\theta'}{(2\pi)^2}
\int d\varepsilon'
\frac{m}{\hbar^2}\left[1-
s'\frac{m\kappa_{\theta'}/\hbar^2}{\sqrt{\left(\frac{m}{\hbar^2}\right)^2
\kappa^2_{\theta'}+\frac{2m\varepsilon'}{\hbar^2}}}\right]\times \\
\nonumber
&& \frac{\pi\hbar^2}{m\tau}\delta(E_{sk}-\varepsilon')
\left[1+ss'\frac{(\alpha^2+\beta^2)\cos(\theta-\theta')+
2\alpha\beta\sin(\theta+\theta')}{\kappa_\theta\kappa_{\theta'}}\right]=
\frac{1}{\tau}[f_s^1(\mathbf{k})+f_s^2(\mathbf{k})]. \\
\label{a1}
\end{eqnarray}
\end{widetext}
The rest terms of $\mathrm{St}[f_s(\mathbf{k})]$ containing $f_s^1$ and
$f_s^2$ given by Eqs.~(\ref{solution1}) and (\ref{solution2}) respectively read
\begin{eqnarray}
\nonumber
&& \sum\limits_{s'}\int\frac{d^2k'}{(2\pi)^2}
w(\mathbf{k}s;\mathbf{k'}s')[-f_{s'}^1(\mathbf{k'})]=
\left[-\frac{\partial f^0(E_{sk})}{\partial E_{sk} }\right] \times  \\
\nonumber
&& \sum\limits_{x,y}
\frac{-s e E_{x,y}}{2\hbar\kappa_\theta}\left(
\frac{\mid\alpha^4-\beta^4\mid}{\alpha^2+\beta^2}
\left\{ \begin{array}{c}
\cos\theta \\ 
\sin\theta
\end{array}\right\}
+ 2\alpha\beta \left\{ \begin{array}{c}
\sin\theta \\ 
\cos\theta
\end{array}\right\}\right. \\
&& \times \left.\left[1+\frac{(\alpha^2+\beta^2)^2}{4\alpha^2\beta^2}\left(
\frac{\mid\alpha^2-\beta^2\mid}{\alpha^2+\beta^2}-1\right)\right] \right),
\label{a2}
\end{eqnarray}
\begin{eqnarray}
\nonumber
&& \sum\limits_{s'}\int\frac{d^2k'}{(2\pi)^2}
w(\mathbf{k}s;\mathbf{k'}s')[-f_{s'}^2(\mathbf{k'})]=
\left[-\frac{\partial f^0(E_{sk})}{\partial E_{sk} }\right]\times  \\
\nonumber
&& \sum\limits_{x,y} 
\frac{s e E_{x,y}}{2\hbar\kappa_\theta}\left[
\cos\theta\left(a_{x,y}+b_{x,y}\frac{\alpha^2+\beta^2-\mid\alpha^2-\beta^2\mid}{2\alpha\beta}
\right)\right. \\
&& + \left. \sin\theta 
\left(b_{x,y}+a_{x,y}\frac{\alpha^2+\beta^2-\mid\alpha^2-\beta^2\mid}{2\alpha\beta}
\right) \right].
\label{a3}
\end{eqnarray}
Substituting Eqs.~(\ref{a1})--(\ref{a3}) into the master equation (\ref{master2})
one can easily establish Eqs.~(\ref{a})--(\ref{b})
for unknown coefficients $a_{x,y}$, $b_{x,y}$.

\subsection{Spin-coherent case}

To prove the solution of Eq.~(\ref{d-h-2}) we take the integrals
in its right hand side in a similar way as before.
Indeed, after the substitution of $f_{11(22)}$ given by Eq.~(\ref{elegant})
and $f_{12}=f_{21}=0$ into the collision term we have
\begin{eqnarray}
\nonumber
&& {\mathrm St}[\hat{f}({\mathbf k})]_{11(22)}=
\frac{f_{11(22)}}{\tau}\mp \sum\limits_{x,y} eE_{x,y}
\left[-\frac{\partial f^0(E_{+(-)k})}{\partial E_{+(-)k} }\right] \\
&& \times\frac{1}{\hbar\kappa_\theta}
\left\{\begin{array}{c}
(\alpha^2+\beta^2)\cos\theta + 2\alpha\beta\sin\theta \\ 
(\alpha^2+\beta^2)\sin\theta + 2\alpha\beta\sin\theta\cos\theta
\end{array}\right\},
\label{a4}
\end{eqnarray}
\begin{eqnarray}
\nonumber
&& {\mathrm St}[\hat{f}({\mathbf k})]_{12(21)}=
\pm\frac{i(\alpha^2-\beta^2)}{2\hbar \kappa_\theta}
\sum\limits_{x,y} eE_{x,y}
\left\{\begin{array}{c} \sin\theta \\ -\cos\theta
\end{array}\right\} \\
&& \times\left[-\frac{\partial f^0(E_{+k})}{\partial E_{+k}}
-\frac{\partial f^0(E_{-k})}{\partial E_{-k} }
\right].
\label{a5}
\end{eqnarray}
Using Eqs.~(\ref{a4})--(\ref{a5}) and the velocity matrix 
elements (\ref{v11x})--(\ref{v12y})
one can easily prove, that the spin-incoherent solution (\ref{elegant})
with $f_{12}=f_{21}=0$ satisfies Eq.~(\ref{d-h-2}).

\end{appendix}

\bibliography{old.bib,new.bib}

\end{document}